\documentclass[aps,preprint,nofootinbib,superscriptaddress]{revtex4-1}
\usepackage{amssymb}

\usepackage{epsfig}
\usepackage[bookmarksnumbered,bookmarksopen,colorlinks,citecolor=blue,linkcolor=blue]{hyperref}
\begin{document}

%\date{\today}

\title{Systematic study of shell gaps in nuclei  }

\author{Qiuhong Mo, Min Liu and Ning Wang}
\thanks{wangning@gxnu.edu.cn}
\address{Department of Physics, Guangxi Normal University,
Guilin 541004, P. R. China}

\begin{abstract}
 The nucleon separation energies and shell gaps in nuclei over the whole nuclear chart are systematically studied with eight global nuclear mass models. For unmeasured neutron-rich and super-heavy regions, the uncertainty of the predictions from these different mass models is still large. The latest version (WS4) of the Weizs\"acker-Skyrme mass formula, in which the isospin dependence of model parameters is introduced into the macroscopic-microscopic approach inspired by the Skyrme energy-density functional, is found to be the most accurate one in the descriptions of nuclear masses, separation energies and shell gaps. Based on the predicted shell gaps in nuclei, the possible magic numbers in super-heavy nuclei region are investigated. In addition to the shell closures at $N=184, Z=114$, the sub-shell closures at around $N=178, Z=120$ could also play a role for the stability of super-heavy nuclei.

\end{abstract}

\maketitle

\begin{center}
\textbf{I. INTRODUCTION}
\end{center}

The development of global nuclear mass formulas is of great importance for not only nuclear physics but also nuclear astrophysics. In nuclear physics, the study of nuclear properties of extremely neutron-rich nuclei and the shell evolution attracted much attention in recent years. In addition, as the necessary theoretical tools, the global nuclear mass formulas can provide some crucial information on the synthesis of super-heavy nuclei \cite{Ogan10,Naza,Sob,Zhou12}, including the shell corrections and neutron separation energies of super-heavy nuclei which are required in the predictions of the survival probability of compound nuclei with a statistical model \cite{Wang11}, the magic numbers around the predicted island of stability, and the $Q$-values in the $\alpha$-decay process etc. \cite{Zhang}. In nuclear astrophysics, a realistic model of an r-process, that would accurately predict the observed elemental abundances, needs a large set of various nuclear characteristics as the input. The essential input is the beta-decay energies which define the rate of evolution along the r-process path, and the neutron-separation energies which determine the position of the r-process path on the nuclear chart. Although a great effort has been devoted in recent decades to accurate measurements of masses of the unstable nuclei, the masses of most nuclei along the r-process path are still unknown and the model predictions for these neutron-rich nuclei play a key role for the study of the r-process \cite{Li12,Xu13,Bret12,Peter12}.

\begin{figure}
\includegraphics[angle=-0,width= 0.9 \textwidth]{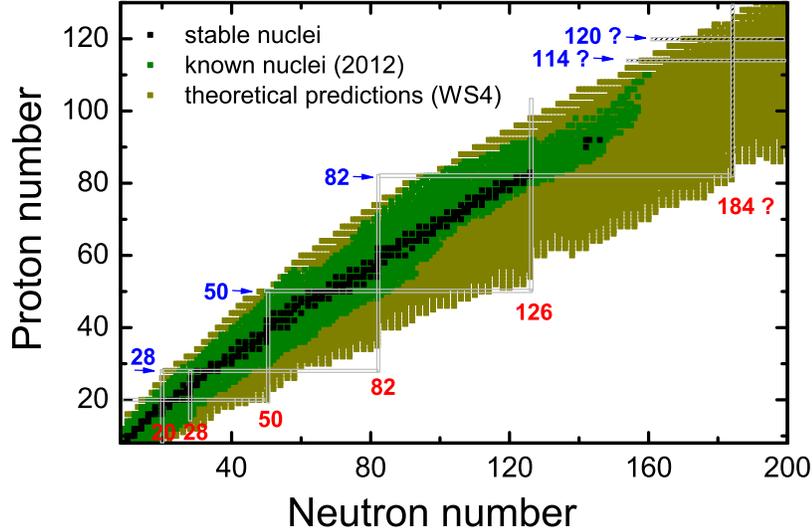}
 \caption{(Color online) Nuclear landscape. The black and green squares denote the stable nuclei and known nuclei in AME2012 \cite{Audi12}, respectively. The dark yellow squares denote the theoretical predictions with the WS4 mass formula \cite{WS4}. The gray solid lines denote the known magic numbers and the dashed lines denote the possible magic numbers in super-heavy region.}
\end{figure}

Available nuclear mass formulas for the predictions of unknown masses include global and local formulas. For the global formulas, the model parameters are usually determined by essentially all measured masses and the masses of almost all bound nuclei can be calculated. Some global nuclear mass models have been successfully
established with an rms errors of about several hundreds keV to one MeV with respect to all measured nuclear masses. These models include: 1) various macroscopic-microscopic mass models such as the finite range droplet model (FRDM) \cite{Moll95}, the Lublin-Strasbourg-Drop (LSD) model \cite{Pom03} and the recently proposed Weizs\"acker-Skyrme (WS) formula \cite{Wang,Wang10,Liu11,WS4}; 2) various microscopic mass models based on the mean-field concept such as the non-relativistic Hartree-Fock-Bogoliubov (HFB) approach with the Skyrme energy-density functional \cite{HFB17, Gor13, HFB27} or the Gogney forces and the relativistic mean-field (RMF) models \cite{Zhao10,Geng05,Meng13}; 3) the Duflo-Zuker (DZ) mass model \cite{DZ28}; and some other global mass models. The local mass formulas are generally based on algebraic or systematic approaches. They predict the masses of unknown nuclei from the masses of known neighboring nuclei, such as the Garvey-Kelson relations \cite{GK}, the isobaric multiplet mass equation \cite{Ormand,Lenzi}, the residual proton-neutron interactions \cite{Zhao1,Zhao12} and the image reconstruction technique (like the CLEAN algorithm \cite{Mora}  and the radial basis function method \cite{WangRBF,Niu13}). The main difficulty of the local mass formulas is that the model errors rapidly increase for nuclei far away from the measured nuclei. In Fig. 1, we show the nuclear landscape. The green and dark yellow squares denote the known nuclei \cite{Audi12} and the predicted ones with the very recent WS4 mass formula \cite{WS4}, respectively. The masses of about 5000 nuclei are still unknown and need the predictions (extrapolation) from the global mass models. It is therefore a great challenge for the global mass models to accurately describe the masses of all nuclei over the whole nuclear chart based on the measured masses of about 2400 nuclei to determine the model parameters. To test the reliability of the global mass formulas, as many as possible mass-related observables should be investigated.

Nucleon separation energies of nuclei not only are intimately related with the particle drip lines, but also provide helpful information on the magic numbers especially the shell closures in the super-heavy region and the shell evolutions in neutron-rich nuclei. As a measure of the discontinuity in the two-neutron separation energy $S_{2n}$ at neutron magic numbers, the shell gap is a sensitive quantity to test the theoretical models \cite{Lun03}. In the WS4 framework, the accuracy of the WS formula is further improved by taking into account the surface diffuseness correction for unstable nuclei. The rms deviation with respect to essentially all the available mass data falls to 298 keV, crossing the 0.3 MeV accuracy threshold for the first time within the mean-field framework. It is therefore interesting and necessary to test the model from the shell gaps in nuclei. In this work, we will systematically investigate the nucleon separation energies and shell gaps in nuclei with eight different global mass models, including the FRDM \cite{Moll95}, HFB17 \cite{HFB17}, HFB27 \cite{HFB27}, DZ28 \cite{DZ28}, WS \cite{Wang}, WS* \cite{Wang10}, WS3 \cite{Liu11} and WS4 \cite{WS4} models.  The paper is organized as follows: In Sec. II, the Weizs\"acker-Skyrme (WS4) mass formula is briefly introduced for the reader's convenience. In Sec. III, the results about the two-neutron separation energies and shell gaps in nuclei will be presented and the comparisons between different models will be discussed. Finally, a summary is given in Sec. IV.

\begin{center}
\textbf{II. WEIZS\"ACKER-SKYRME MASS FORMULA}
\end{center}

In the Weizs\"acker-Skyrme (WS4) mass formula \cite{WS4}, the total energy of a nucleus is written as
a sum of the liquid-drop energy, the Strutinsky shell correction and the residual correction,
\begin{eqnarray}
E (A,Z,\beta)=E_{\rm LD}(A,Z) \prod_{k \ge 2} \left (1+b_k
\beta_k^2 \right )+\Delta E (A,Z,\beta) + \Delta_{\rm res}.
\end{eqnarray}
The liquid-drop energy of a spherical nucleus $E_{\rm LD}(A,Z)$ is
described by a modified Bethe-Weizs\"acker mass formula,
\begin{eqnarray}
E_{\rm LD}(A,Z)=a_{v} A + a_{s} A^{2/3}+ E_C + a_{\rm sym} I^2 A f_{s} +
a_{\rm pair}  A^{-1/3}\delta_{np} + \Delta_W,
\end{eqnarray}
with the isospin asymmetry $I=(N-Z)/A$.
\begin{eqnarray}
E_C=a_c \frac{Z^2}{A^{1/3}} \left ( 1- 0.76 Z^{-2/3} \right)
\end{eqnarray}
and
\begin{eqnarray}
a_{\rm sym}=c_{\rm sym} \left [1-\frac{\kappa}{A^{1/3}}+ \xi  \frac{2-|I|}{ 2+|I|A} \right ]
\end{eqnarray}
denote the Coulomb energy term and the symmetry energy coefficient of finite nuclei, respectively.
\begin{eqnarray}
f_s=1+\kappa_s \varepsilon A^{1/3}
\end{eqnarray}
is a correction factor to the symmetry energy considering the surface diffuseness effect of unstable nuclei. Here, $\varepsilon=(I-I_0)^2-I^4$ with $I_0=0.4A/(A+200)$ being the isospin asymmetry of the nuclei along the $\beta$-stability line described by Green's formula. The $a_{\rm pair}$ term empirically describes the odd-even staggering effect. In WS4, the $I^2$ term in the isospin dependence of $\delta_{np}$ is further introduced for a better description of the masses of even-A nuclei:
\begin{eqnarray}
\delta_{np}= \left\{
\begin{array} {l@{\quad:\quad}l}
  (2 - |I|-I^2)\frac{17}{16}  &   N {\rm ~and~} Z {\rm ~even }    \\
      |I|-I^2  &   N {\rm ~and~} Z {\rm ~ odd }    \\
  1 - |I|  &   N {\rm ~even,~} Z {\rm ~odd,~ } {\rm ~and~ } N>Z   \\
  1 - |I|  &   N {\rm ~odd,~} Z {\rm ~even,~ } {\rm ~and~ } N<Z   \\
  1             &   N {\rm ~even,~} Z {\rm ~odd,~ } {\rm ~and~ } N<Z   \\
  1             &   N {\rm ~odd,~} Z {\rm ~even,~ } {\rm ~and~ } N>Z   \\
\end{array} \right.
\end{eqnarray}
$\Delta_W$ in Eq.(2) denotes the Wigner correction term for heavy nuclei \cite{Liu11}. The dependence of the macroscopic energy on the nuclear deformations in the WS formula is given by an analytical expression $E_{\rm LD} \prod  \left (1+b_k \beta_k^2 \right )$ for nuclei with small deformations. The curvatures of the parabolas $b_k$ are written as ,
\begin{eqnarray}
b_k=\left ( \frac{k}{2} \right ) g_1A^{1/3}+\left ( \frac{k}{2} \right )^2 g_2 A^{-1/3}.
\end{eqnarray}
Here, the mass dependence of the curvature is obtained from the Skyrme energy-density functional in which the influence of nuclear surface diffuseness and symmetry energy on the deformation energies of nuclei is self-consistently involved.

The microscopic shell correction is expressed as
\begin{eqnarray}
\Delta E=c_1 f_d E_{\rm sh} + |I| E_{\rm sh}^{\prime}.
\end{eqnarray}
Here, $c_1$ is a scale factor. $f_d$ denotes the corresponding correction due to the surface diffuseness,
\begin{eqnarray}
 f_d=1+\kappa_d \varepsilon.
\end{eqnarray}
$E_{\rm sh}$ and $ E_{\rm sh}^{\prime}$ denote the shell energy of a nucleus and of its mirror nucleus obtained with the traditional Strutinsky procedure by setting the smoothing parameter $\gamma=1.2\hbar\omega_0$ and the order $p=6$ of the Gauss-Hermite polynomials. The $|I|$ term in $\Delta E$ is to take into account the mirror effect \cite{Wang10} from the isospin symmetry.

\begin{table}
\caption{ Model parameters of the mass formula WS4. }
\begin{tabular}{cccc }
\hline\hline
  Parameter                         & ~~~~~~Value~~~~~~& ~~~ Parameter~~~       & ~~~Value~~~\\ \hline
 $a_v  \; $ (MeV)                  &   $-15.5181$   & $g_1 $                            &   0.01046 \\
 $a_s \; $  (MeV)                  &   17.4090      &  $g_2 $                           &   $-0.5069$ \\
 $a_c \; $ (MeV)                   &   0.7092       &  $V_0$ (MeV)                      &   $-45.8564$ \\
 $c_{\rm sym} $(MeV)               &   30.1594      &   $r_0$ (fm)                      &   1.3804   \\
 $\kappa \;  $                     &   1.5189       &   $a_0 $ (fm)                       &   0.7642   \\
 $\xi \;  $                        &   1.2230       &  $\lambda_0$                      &   26.4796  \\
  $a_{\rm pair} $(MeV)             &   $-5.8166$    &  $c_1  \; $                       &   0.6309   \\
  $c_{\rm w} $ (MeV)               &   0.8705       &  $c_2  \; ({\rm MeV} ^{-1})$      &   1.3371  \\
  $\kappa_{s} $                    &   0.1536       &  $\kappa_{d} $                    &   5.0086  \\
   \hline\hline
\end{tabular}
\end{table}

\begin{table}
\caption{ Comparison of the correction terms adopted in the WS series models. "$+$" and "$-$" denote with and without the corresponding terms being taken into account, respectively. Here, the deformations of nuclei involved in the calculations are also listed for comparison. }
\begin{tabular}{llccc }
\hline\hline
  Version~~~      &  Deformations~~~ &   ~~~$|I| E_{\rm sh}^{\prime} $ term ~~~  & ~~~$\Delta_{\rm res}$ term~~~ & ~~~$\varepsilon$ terms \\ \hline
  WS       &    $\beta_2$, $\beta_4$              & $ - $                           &  $ - $              &$ - $            \\
  WS*      &    $\beta_2$, $\beta_4$,  $\beta_6$  & $ + $                           &  $ - $              &$ - $            \\
  WS3      &    $\beta_2$, $\beta_4$,  $\beta_6$  & $ + $                           &  $ + $              &$ - $            \\
  WS4      &    $\beta_2$, $\beta_4$,  $\beta_6$  & $ + $                           &  $ + $              &$ + $            \\
   \hline\hline
\end{tabular}
\end{table}

The optimal values of the 18 independent model parameters listed in Table I are obtained based on the 2353 ($N$ and $Z\ge8$) measured nuclear masses $M_{\rm exp}$ in AME2012 and searching for the minimal rms deviation with respect to the masses $ \sigma^2=  \frac{1}{m}\sum [M_{\rm exp}^{(i)}-M_{\rm th}^{(i)}]^2 $. The rms deviation with respect to essentially all the available mass data falls to 298 keV with the WS4 formula, the best value ever found within the mean-field framework. In Table II, we also list the main difference in the four versions of the WS series models. Based on the WS model, the mirror effect is further considered in the version WS*. The residual correction term $\Delta_{\rm res}$ \cite{Liu11} and the surface diffuseness correction $\varepsilon$ terms are further involved in the version WS3 and WS4, respectively.

\begin{center}
\textbf{III. RESULTS AND DISCUSSIONS}
\end{center}

In this sections, we first systematically investigate the two-neutron separation energies of nuclei. Then, the shell gaps including the proton and neutron shell gaps will be studied with the eight global mass models.

\begin{center}
\textbf{A. Two-neutron separation energies}
\end{center}

\begin{figure}
\includegraphics[angle=-0,width= 0.8\textwidth]{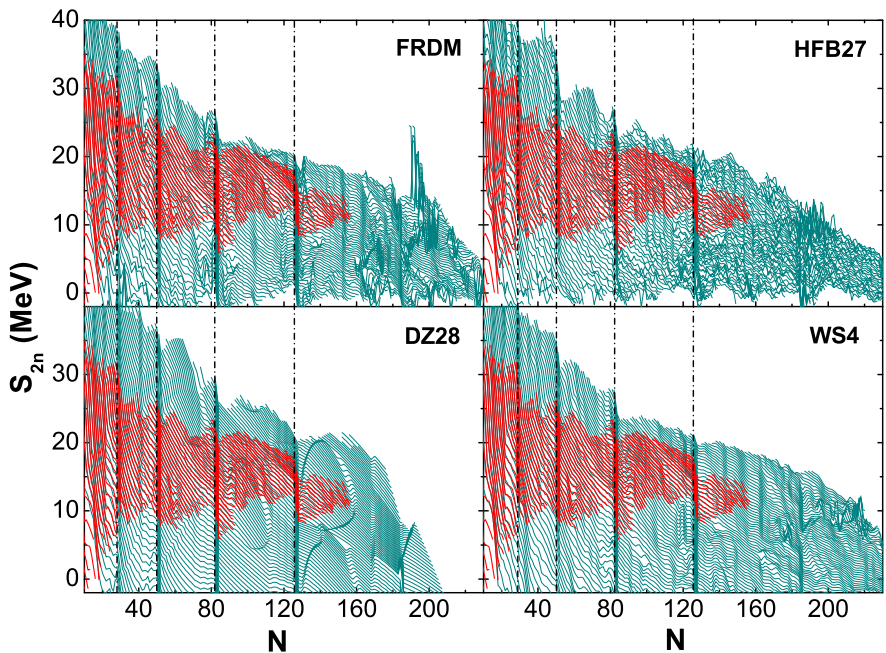}
 \caption{(Color online) Two-neutron separation energy $S_{\rm 2n}$ as a function of neutron number $N$. The red and dark cyan curves denote the experimental data and the model predictions, respectively. The dot-dashed lines denote the known neutron magic numbers.
}
\end{figure}

Based on the binding energies $B(N,Z)$ of nuclei with neutron number $N$ and charge number $Z$, one can obtain the corresponding two-neutron separation energy,
\begin{eqnarray}
S_{\rm 2n} (N,Z,)=B(N,Z)-B(N-2,Z).
\end{eqnarray}
The rms deviations with respect to the 2123 measured $S_{\rm 2n}$ from the FRDM, HFB27, DZ28 and WS4 models are 0.493, 0.425, 0.336 and 0.276 MeV, respectively. To see the global behavior of $S_{\rm 2n}$, we show in Fig. 2 the surface of two-neutron separation energy obtained from these four mass models. The red and dark cyan curves denote the experimental data based on the measured masses in AME2012 and the model predictions, respectively. The dot-dashed lines indicate the known neutron magic numbers. Each curve denotes the corresponding $S_{\rm 2n}$ of an isotopic chain. One sees that: (1) The two-neutron separation energies of nuclei generally decrease with neutron numbers; (2) The sudden decrease in the $S_{\rm 2n}$ at the magic numbers is evident, which reflects the existence of well-known magic numbers; (3) At the region $N \approx 200$, the fluctuations of $S_{\rm 2n}$ are large for FRDM and HFB27.

\begin{figure}
\includegraphics[angle=-0,width= 1\textwidth]{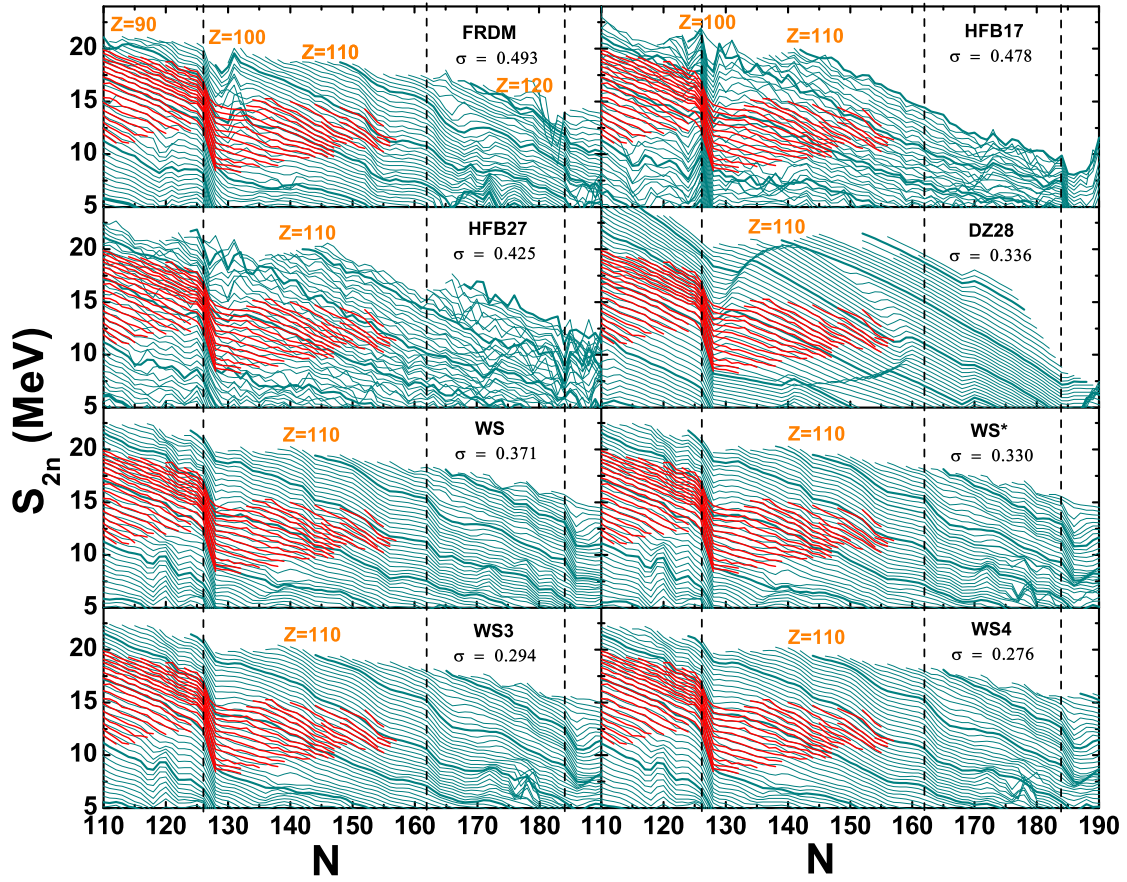}
 \caption{ (Color online) Two-neutron separation energy $S_{\rm 2n}$ from different nuclear mass models. The calculated results for the isotopes with $Z=120$, 110, 100, 90, 80, 70 and 60 are indicated by the thick curves. The value of $\sigma$ (in MeV) denotes the corresponding rms deviations with respect to the 2123 measured $S_{\rm 2n}$.  }
\end{figure}

To see the two-neutron separation energies of nuclei at super-heavy region more clearly, the comparison of the $S_{\rm 2n}$ from eight different mass models is shown in Fig. 3. From the figure, one finds that the fluctuations in $S_{\rm 2n}$ are relatively large from the microscopic HFB calculations. The dashed lines indicate the positions of $N=126$, 162 and 184. The sadden decrease of $S_{\rm 2n}$ calculated with the macroscopic-microscopic models including the FRDM and the WS series models indicates that $N=162$ could be a possible neutron magic number \cite{Sob,Poen06,Zhang}. For the FRDM, the evident peaks in $S_{\rm 2n}$ for the isotopes $87\le Z \le101$ can be observed at $N=132$. To check the behavior of $S_{\rm 2n}$ at $N=132$, the $S_{\rm 2n}$ of Ra isotopes are presented in Fig. 4. The experimental data do not indicate the appearance of an evident sub-shell at $N=132$ in the Ra isotopes, which implies that the single-particle potential adopted in the FRDM should be refined. Comparing with the FRDM, the two-neutron separation energies in Ra isotopes can be much better reproduced with the WS4 formula, which is probably due to the isospin dependence of the potential parameters being taken into account in the WS4 formula. In addition to the isospin-dependence of model parameters, some advantages of the WS4 over the other considered models may come from the fact that it was adjusted to more recent and more neutron-rich experimental data. Comparing with the results of HFB17, the latest HFB27 model gives better results for the Ra isotopes through re-adjusting the model parameters based on more measured masses.

\begin{figure}
\includegraphics[angle=-0,width= 0.7\textwidth]{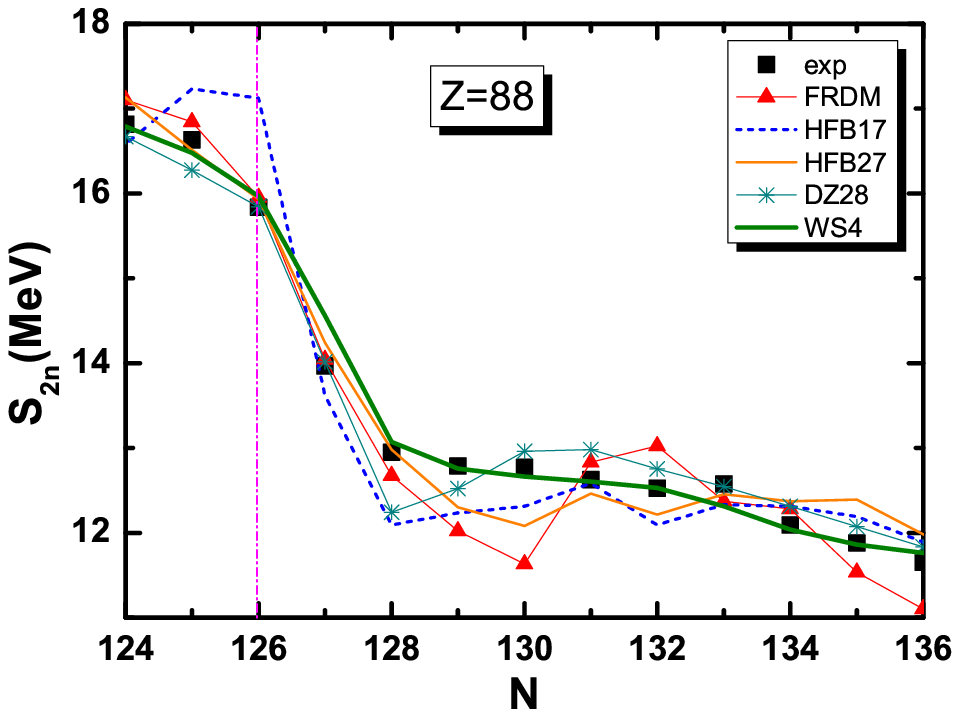}
 \caption{(Color online) $S_{\rm 2n}$ of Ra isotopes as a function of neutron number. The black squares denote the experimental data and the curves denote the predictions from five different models. }
\end{figure}

\begin{center}
\textbf{B. Shell gaps in nuclei}
\end{center}

\begin{table}
\centering
\caption{Rms deviations between data and predictions from eight global mass models (in keV).}
\begin{tabular}{lccccccccc}
 \hline\hline
Rms~~ & ~~$N_{\rm nucl}$~~ & ~~FRDM~~ & ~~HFB17~~ & ~~HFB27~~ & ~~DZ28~~& ~~WS~~ & ~~WS*~~ & ~~WS3~~ & ~~WS4~~\\
\hline
$\sigma (M)$   & $2353$ & $654$ & $576$ & $512$ & $394$ & $525$ & $439$ & $334$ & $298$  \\
$\sigma (S_n)$ & $2199$ & $376$ & $500$ & $425$ & $296$ & $331$ & $316$ & $273$ & $258$ \\
$\sigma (S_{2n})$& $2123$ & $493$ & $478$ & $425$ & $336$ & $371$ & $330$ & $294$ & $276$  \\
$\sigma (S_p)$ & $2150$ & $395$ & $502$ & $434$ & $304$ & $352$ & $333$ & $296$ & $274$  \\
$\sigma (S_{2p})$& $2032$ & $502$ & $524$ & $449$ & $366$ & $431$ & $392$ & $354$ & $322$  \\
$\sigma (\Delta)$&$1689$& $1053$ & $1204$ & $998$ & $789$ & $851$ & $816$ & $762$ & $725$  \\
\hline
\end{tabular}
\end{table}

In this work, the empirical shell gaps in nuclei are defined as the sum of the neutron and proton shell gaps based on the difference of the binding energies,
\begin{eqnarray}
\Delta (N,Z)=\Delta_n (N,Z)+\Delta_p (N,Z),
\end{eqnarray}
with
\begin{eqnarray}
\Delta_n (N,Z)=B(N+2,Z)+B(N-2,Z)-2B(N,Z)
\end{eqnarray}
and
\begin{eqnarray}
\Delta_p (N,Z)=B(N,Z+2)+B(N,Z-2)-2B(N,Z).
\end{eqnarray}
In Table III, we list the rms deviations of different mass models with respect to the measured masses, nucleon separation energies and shell gaps.  The line $\sigma(M)$ refers to all the 2353 measured masses ($N\ge 8$, $Z\ge8$) in AME2012, the line $\sigma (S_n)$ to all the 2199 neutron separation energies $S_n$, the line $\sigma (S_{2n})$ to all the 2123 measured $S_{2n}$, the line $\sigma (S_p)$ to the proton separation energies, the line $\sigma (S_{2p})$ to the two-proton separation energies, and the last line $\sigma (\Delta)$ to the 1689 measured shell gaps. For these mass models, the rms deviations with respect to the measured masses are about 600 keV to 300 keV. For the descriptions of the neutron and proton separation energies $S_n$ and $S_p$, the rms errors of the microscopic Skyrme HFB models are relatively larger than those of the other models listed. In addition, we note that the rms deviation with respect to the proton separation energies is larger than that to the neutron separation energies, especially for the WS series models, which implies that some physics related to the protons in nuclei could be still missing in these mass models.

\begin{figure}
\includegraphics[angle=-0,width= 1\textwidth]{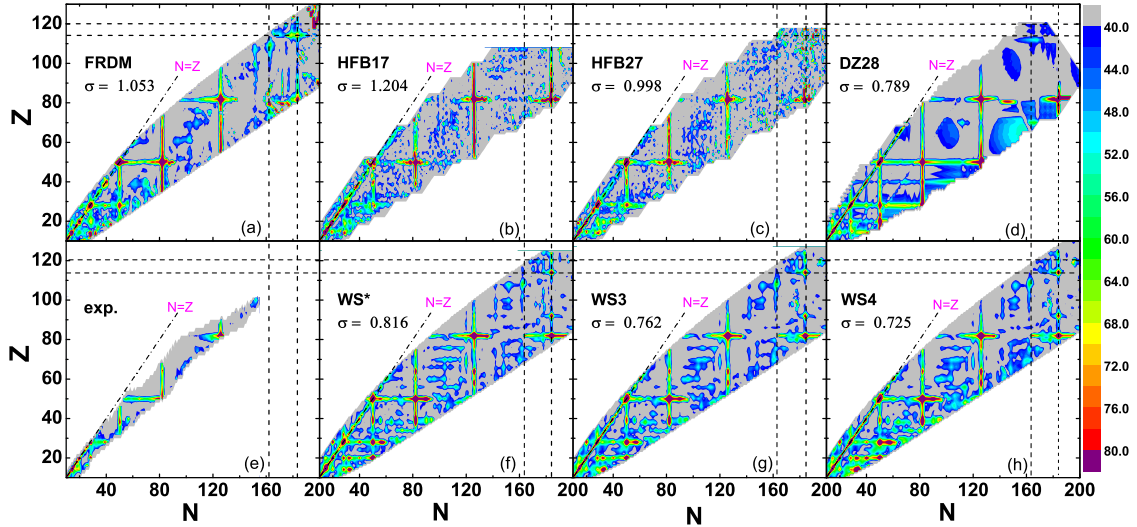}
 \caption{(Color online) Contour plot of shell gaps in nuclei scaled by $A^{1/2}$ from the experimental data and model predictions. The dashed lines indicate the possible magic numbers $N=162,184$ and $Z=114,120$. The dot-dashed line indicates the positions of nuclei with $N=Z$.}
\end{figure}

To explore the global trend of the shell gaps in nuclei, we show in Fig. 5 the contour plot of $\Delta (N,Z) A^{1/2}$ for nuclei over the whole nuclear chart. Here, the shell gap is multiplied by a factor $A^{1/2}$ in order to show the change of $\Delta (N,Z)$ with the same scale for both light and heavy nuclei. From the figure, one can see that the values of $\Delta A^{1/2}$ are significantly larger for nuclei with well-known magic numbers than those of open shell nuclei due to the shell effects. In addition, for nuclei along the $N=Z$ line, the values of $\Delta A^{1/2}$ are also large due to the Wigner effect. For super-heavy nuclei, the values of the shell gaps in nuclei with $N=184$ are relatively larger than those of their neighboring nuclei from all these different mass models. In addition to the evident shell gaps in nuclei with known magic numbers, the sub-shell closures can also be observed from the experimental date in Fig. 5(e). The shell gaps in nuclei with sub-shell closure such as some nuclei with $Z=40$, 70, 76 and those with $N=108$, 152, 162 are also evident from the macroscopic-microscopic calculations. The shell gap could be an effective probe to investigate the fine structure of nuclei caused by the residual shell effects, since the smooth macroscopic part in the nuclear binding energy is removed through the mass difference. For the Skyrme HFB calculations, the large fluctuations in the two-neutron separation energies (see Fig. 3) result in some difficulties to make a clear distinction between the sub-shell closures and fluctuations. The rms deviations (in MeV) with respect to the measured shell gaps $\Delta (N,Z)$ are also presented in the figure for the seven different mass models. The value of $\sigma(\Delta)$ varies from 1.204 MeV of HFB17 to 0.725 MeV of WS4. The rms deviation $\sigma(\Delta)$ is generally larger than the corresponding value of $\sigma(S_{2p})$ by a factor of two for a certain mass model, which is due to that the shell gap is defined as $\Delta=\Delta_p +\Delta_n$ and the model errors from both $S_{2p}$ and $S_{2n}$ affect the results.

\begin{figure}
\includegraphics[angle=-0,width= 0.85\textwidth]{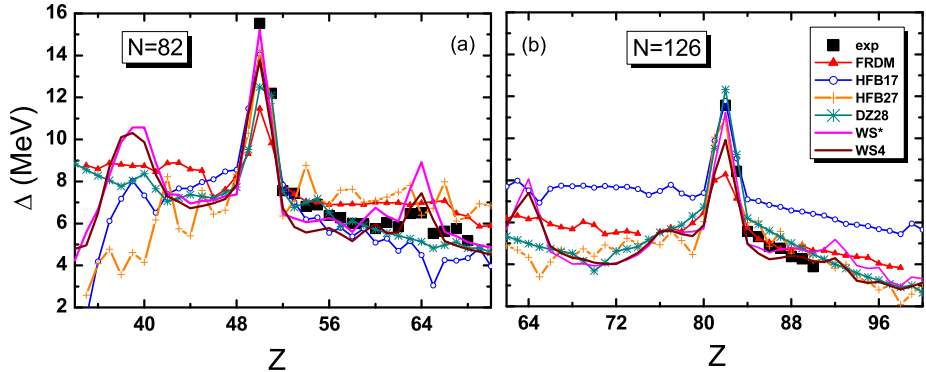}
 \caption{(Color online) Shell gaps for nuclei with $N=82$ and 126. The black squares denote the experimental data, and the curves denote the predictions from different models.}
\end{figure}

\begin{figure}
\includegraphics[angle=-0,width= 0.85\textwidth]{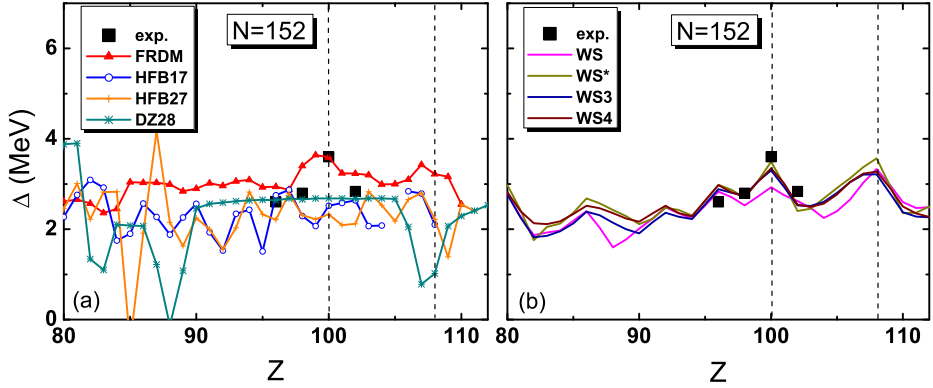}
 \caption{(Color online) The same as Fig. 6, but for nuclei with $N=152$. The dashed lines indicate the positions of $Z=100$ and $Z=108$.}
\end{figure}

In Fig. 6, we compare the different mass models for the descriptions of the shell gaps in nuclei with magic number $N=82$ and 126. The black squares denote the experimental data.  The peak in the experimental data at $Z=64$ as sub-shell closure can be reasonably well reproduced only by the WS series models. For the extremely neutron-rich nuclei around $^{122}$Zr, the deviations of the shell gaps from these different models are very large. The shell gaps in nuclei around $Z=88, N=126$ from the HFB17 calculations are systematically larger than the experimental data by about 2 MeV.

To further check these global mass models for the description of sub-shell closure, we show the predictions of the shell gaps in nuclei with $N=152$ from the eight models in Fig. 7. The calculated cranked Nilsson levels \cite{Zhang12} and the single-particle levels near the Fermi surface from the Woods-Saxon potential \cite{Chas77} and those determined from the experimental information suggest that there exist a proton gap at $Z =100$ and a neutron gap at $N = 152$. From Fig. 7 (a), one sees that the uncertainties of the predicted shell gaps are quite large for the nuclei with $N=152$ from the four different models (FRDM, HFB17, HFB27 and DZ28) and the peak in the experimental data at $Z=100$ can not be distinctly reproduced. Whereas, the results from the WS series models in Fig. 7(b) are generally consistent with each other except the result of the WS model in which the mirror effect is not taken into account, and the sub-shell closure at $Z=100$ can be evidently observed. In addition, the shell closure at $Z=108$ can also be observed with the WS series models.

\begin{figure}
\includegraphics[angle=-0,width= 0.85\textwidth]{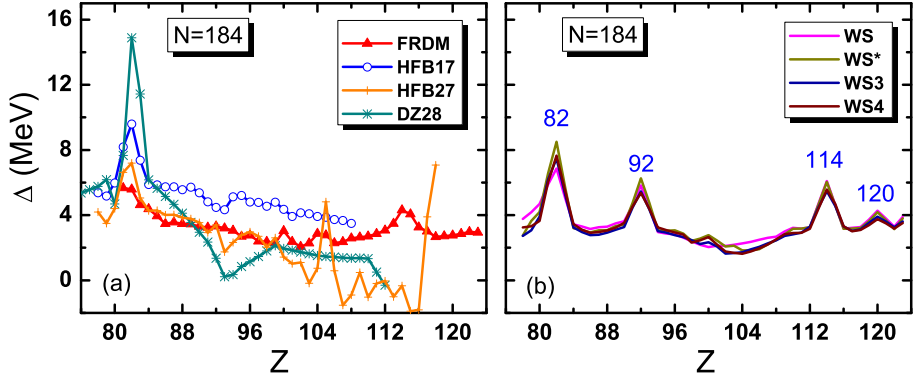}
 \caption{(Color online) The same as Fig. 7, but for super-heavy nuclei with $N=184$.}
\end{figure}

The same as Fig. 7, we show in Fig. 8 the predictions of the shell gaps in nuclei with $N=184$ from the eight nuclear mass models. From Fig. 8 (a), one sees that the uncertainties of the predicted shell gaps are also quite large for the super-heavy nuclei with $N=184$ from the four different models. Whereas, the results from the WS series models in Fig. 8(b) are highly consistent with each other due to the similar theoretical framework adopted. According to the calculations of the WS series models, the sub-shell closures at $Z=92$ and 120 can also be observed in addition to the two evident magic numbers $Z=82$ and 114. We also note that the sub-shell closure at $Z=92$ can also be evidently observed from the relativistic mean field calculations \cite{Geng05} and the measured relatively large shell gap in nuclei around $^{234}$U ($N=142$, $Z=92$) in Fig. 5(e).

\begin{figure}
\includegraphics[angle=-0,width= 1\textwidth]{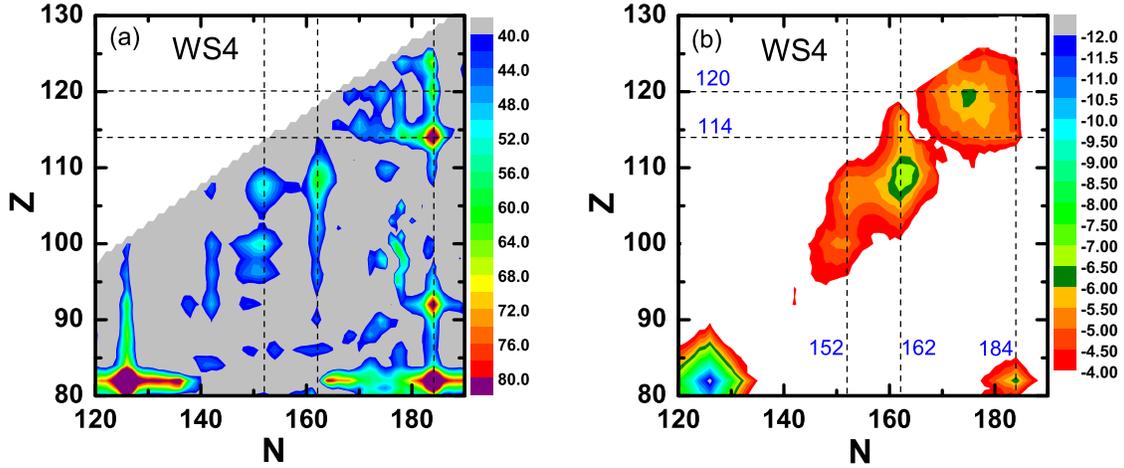}
 \caption{(Color online) (a) Shell gaps of super-heavy nuclei with the WS4 formula. (b) Corresponding shell corrections for nuclei in (a) with the same mass formula. The dashed lines indicate the positions of $N=152$, 162, 184, and $Z=114$, 120.}
\end{figure}

To investigate the possible magic numbers in super-heavy nuclei, the shell corrections are simultaneously studied in addition to the shell gaps. In Fig. 9, the shell gaps and shell corrections of nuclei from the WS4 calculations are compared. The dashed lines indicate the possible magic numbers. For the known doubly-magic nucleus $^{208}$Pb and the deformed doubly-magic nucleus $^{270}$Hs ($N=162$, $Z=108$)\cite{Poen06,Zhang,Sob}, both the shell gaps and the shell corrections (in absolute value) are significantly larger than the corresponding values of their neighboring nuclei. For the super-heavy nucleus $^{298}$Fl ($N=184, Z=114$), however, the result from the shell gap is not well consistent with that from the shell correction. Due to the influence of the sub-shell closures at $N=178$ and $Z=120$, the shell correction of nucleus $^{296}$120 with a value of 6.2 MeV is larger than that of $^{298}$Fl by one MeV.

\begin{center}
\textbf{IV. SUMMARY}
\end{center}

Based on eight different global nuclear mass models with accuracy at the level of 600 keV to 300 keV, the two-neutron separation energies and shell gaps in nuclei have been systematically investigated. We find that:
\begin{itemize}
\item  The sudden decrease in the two-neutron separation energies $S_{\rm 2n}$ at the known magic numbers is evident according to the predictions of all the eight mass models.
\item  The fluctuations in $S_{\rm 2n}$ are relatively large from the microscopic Skyrme HFB calculations, and the fluctuations at the region $N \approx 200$ from the FRDM are also large.
\item  The nuclei with sub-shell closure can be well identified in addition to the known magic nuclei based on the extracted shell gaps from the data and the predictions from the macroscopic-microscopic mass models.
\item  The observed sub-shell closure at $Z=64$ and $Z=100$ from the experimental data can be reasonably well reproduced only by the WS series models, and the experimental data do not indicate the appearance of an evident sub-shell at $N = 132$ in the Ra isotopes.
\item The rms deviation with respect to the proton separation energies is larger than that to the corresponding neutron separation energies, especially for the WS series models, which might imply that some physics related to the protons in nuclei could be still missing in these global mass models.
\item  All eight models predict that $N=184$ is a neutron magic number from $S_{2n}$ and the shell gaps. According to the calculations of the WS series models for super-heavy nuclei with $N=184$, the sub-shell closures at $Z = 92$ and 120 can also be observed in addition to the two evident magic numbers $Z = 82$ and 114.
\item  The shell closures in $^{208}$Pb and the deformed doubly-magic nucleus $^{270}$Hs can be unambiguously observed from  both the shell gaps and the shell corrections, whereas for the nucleus $^{298}$Fl with possible shell closure according to the large value of the shell gap, the corresponding shell correction is smaller than that of its neighboring nucleus $^{296}$120 by one MeV due to the influence of sub-shell closures at $N=178$ and $Z=120$ according to the predictions of the WS4 formula.
\end{itemize}

\begin{center}
\textbf{ACKNOWLEDGEMENTS}
\end{center}

This work was supported by National Natural Science Foundation of China, Nos 11275052, 11365005, 11365004, and 11422548. The nuclear mass tables with the WS formulas are available from http://www.imqmd.com/mass/.

\end{document}